\documentclass[twocolumn,preprintnumbers,amsmath,amssymb]{revtex4-1}
\usepackage{graphicx}
\usepackage[svgnames]{xcolor}
\usepackage{bm}
\usepackage{multirow}
\usepackage{hyperref}

\setcitestyle{super} 

\newif\ifshowcomments\showcommentstrue

\begin{document}

\title{Anomalous transverse response of Co$_2$MnGa and universality of the room-temperature $\alpha^A_{ij}/\sigma^A_{ij}$ ratio across topological magnets}

\author{Liangcai Xu$^{1,2}$, Xiaokang Li$^{1,2}$, Linchao Ding$^{1}$,  Taishi Chen$^{3,4,5}$, Akito Sakai$^{3,4,5}$,  Beno\^{\i}t Fauqu\'e$^{6}$, Satoru Nakatsuji$^{3,4,5}$, Zengwei Zhu$^{1,*}$ and Kamran Behnia$^{2,*}$}
\affiliation{(1) Wuhan National High Magnetic Field Center and School of Physics, Huazhong University of Science and Technology,Wuhan 430074, China\\
 (2)Laboratoire de Physique Et Etude des Mat\'{e}riaux (UPMC-CNRS), ESPCI Paris, PSL Research University,75005 Paris, France\\
(3)Institute for Solid State Physics, University of Tokyo, Kashiwa,Chiba 277-8581, Japan\\
(4)Department of Physics, University of Tokyo, Hongo, Bunkyo-ku, Tokyo 113-0033, Japan\\
(5)CREST, Japan Science and Technology Agency (JST), 4-1-8 Honcho Kawaguchi, Saitama 332-0012, Japan\\
(6) Coll\`ege de France, 11 place Marcelin Berthelot, 75005 Paris, France\\
}
\date{\today}
\begin{abstract}
The off-diagonal (electric, thermal and thermoelectric) transport coefficients of a solid can acquire an anomalous component due to the non-trivial topology of the Bloch waves. We present a study of the anomalous Hall (AHE), Nernst (ANE) and thermal Hall effects (ATHE) in the Heusler Weyl ferromagnet Co$_2$MnGa. The Anomalous Wiedemann-Franz law, linking electric and thermal responses, was found to be valid over the whole temperature window. This indicates that the AHE has an intrinsic origin and the Berry spectrum is smooth in the immediate vicinity of the Fermi level. From the ANE data, we extract the magnitude and temperature dependence of $\alpha^A_{ij}$ and put under scrutiny the $\alpha^A_{ij}/\sigma^A_{ij}$ ratio, which approaches k$_B$/e  at room temperature. We show that in various topological magnets the room-temperature magnitude of this ratio is a sizeable fraction of k$_B$/e and argue that the two anomalous transverse coefficients depend on universal constants, the Berry curvature averaged over a window set by either the Fermi wavelength (for Hall) or the de Broglie thermal length (for Nernst). Since the ratio of the latter two is close to unity at room temperature, such a universal scaling finds a natural explanation in the intrinsic picture of anomalous transverse coefficients.
\end{abstract}
\maketitle

The  anomalous Hall effect (AHE)~\cite{Nagaosa2010,Xiao2010,Nakatsuji2015} has thermoelectric and thermal counterparts, which emerge whenever the longitudinal electric field is replaced by a longitudinal temperature gradient. When intrinsic, the anomalous Nernst (ANE) and the anomalous Righi-Leduc (or thermal Hall (ATHE) effects, like AHE, are caused by the non-vanishing Berry curvature of Bloch waves in the host solid~\cite{Xiao2006,Onoda2008,Li2017,Ikhlas2017,Sakai2018}. A recent theme of interest is the magnitude of these anomalous coefficients. Universal scaling between the amplitudes of magnetization and the amplitude of the transverse response has been sought and invalidated~\cite{Xiao2006,Ikhlas2017,Guin2019}.  Correlations among anomalous transverse coefficients themselves has also been explored. The anomalous version of the Wiedemenn-Franz law, establishing a link between the magnitude of AHE and ARLE has been tested~\cite{Li2017,Xu2018,Sugii2019} and the validity of the Mott's relation linking AHE and ANE has been investigated~\cite{Li2017,Ikhlas2017,Wuttke2019,Ding2019,Park2019}.

In this paper, we present an extensive study of the three anomalous transport coefficients of Co$_2$MnGa and report on several new findings. First of all, we find that the anomalous Wiedemann-Franz law (WF) holds in this system between 2K and 300 K. To the best of our knowledge, this is the unique case of such an extensive verification. Second, we quantify $\alpha^A_{ij}$ and confirm that, as  noticed previously\cite{Sakai2018,Guin2019}, $\alpha^A_{ij}$ is exceptionally large in this system. We quantify the temperature dependence of $\alpha^A_{ij}/\sigma^A_{ij}$ ratio and find that it tends to saturate to a sizeable fraction of k$_B$/e at high temperature. We then show that ratio of $\alpha^A_{ij}/\sigma^A_{ij}$ at room temperature in all known topological magnets lies between 0.2 k$_B$/e and  0.9 k$_B$/e in spite of a tenfold variation in the amplitude of $\alpha^A_{ij}$. We will see below that such a striking correlation is expected when the anomalous transverse response is intrinsic. It implies that the large $\alpha^A_{ij}$ in this system ~\cite{Sakai2018,Guin2019} is unsurprising given its large $\sigma^A_{ij}$.


Co$_2$MnGa is a Weyl ferromagnet\cite{Belopolski2019,Sakai2018} with a Heusler L2$_1$ structure. It has  a Curie temperature of T$_C$ $\approx$ 694 K and a saturation moment of 4 $\mu_B$ per formula unit~\cite{Webster1971}. Previous studies have reported on strain-induced magnetic anisotropy~\cite{Pechan2005} and negative anisotropic magnetoresistance~\cite{Sato2018}. Recently, a large anomalous Hall conductivity exceeding 1000 $\Omega$ cm$^{-1}$ has been observed in thin films and more 2000 $\Omega$ cm$^{-1}$ in crystals of Co$_2$MnGa~\cite{Manna2018,Markou2019,Sakai2018}. A large anomalous Nernst effect was also reported by a number of previous authors~\cite{Sakai2018,Reichlova2018,Guin2019,Park2019}. The magnitude of room-temperature ANE in crystals was found to be remarkably large~\cite{Sakai2018,Guin2019}. Guin \textit{et al.}~\cite{Guin2019} reported that S$^A_{xy}$ is much larger than what is expected according to a simple scaling with magnetization and Sakai \textit{et al.} invoked a quantum critical scaling for $\alpha^A_{xy}$~\cite{Sakai2018}. In this context, and as we will see below, a comparison of the  $\alpha^A_{ij}/\sigma^A_{ij}$ across various magnets is illuminating. Despite the fact that $\sigma^A_{ij}$ varies hundredfold (8  to 1000 $\Omega^{-1}$cm$^{-1}$) in different topological magnets, the ratio of $\alpha^A_{ij}/\sigma^A_{ij}$ remains a sizeable fraction of k$_B$/e.

\begin{figure}
\includegraphics[width=8.5cm]{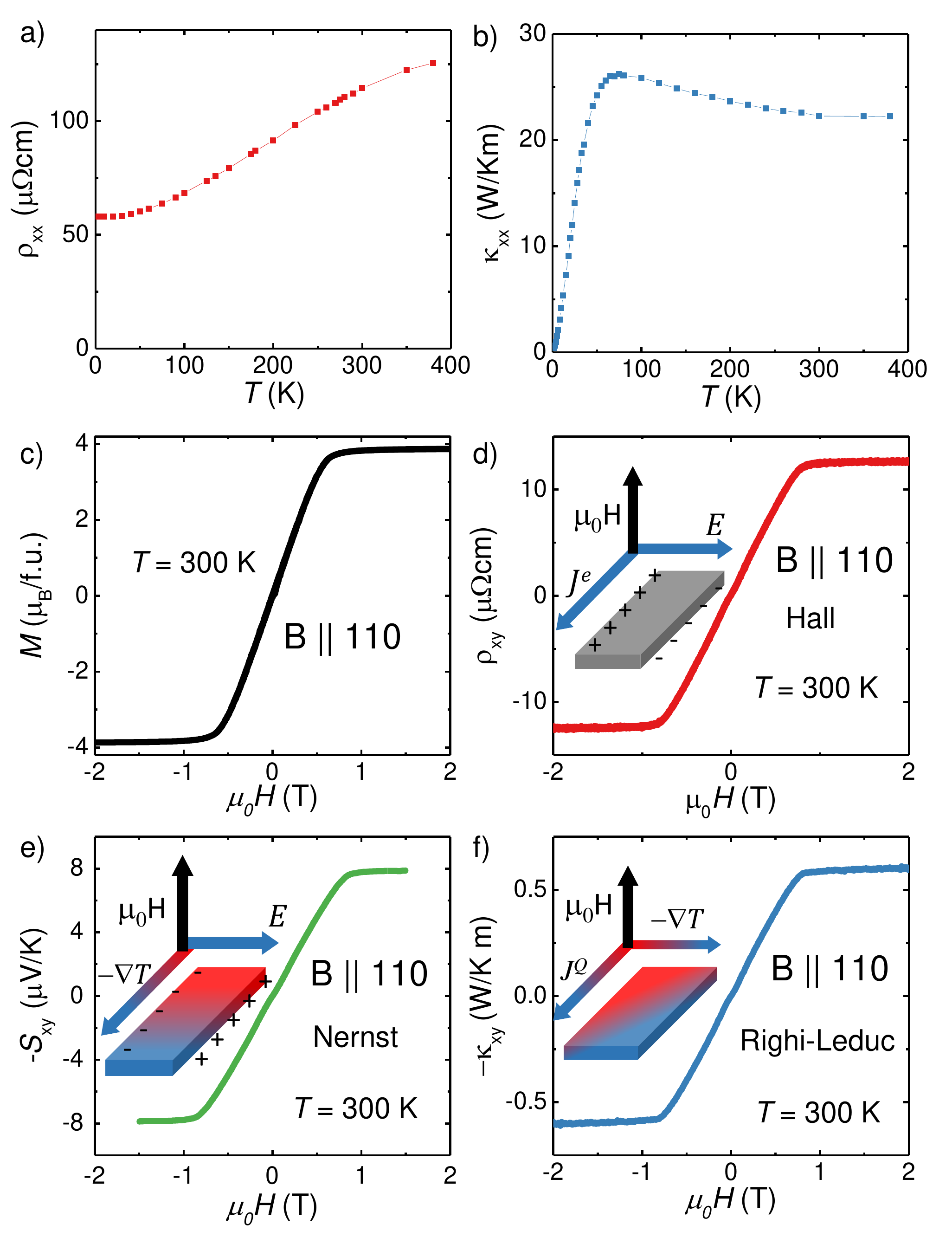}
\caption{\textbf{Basic transport and anomalous transverse responses at (\textit{T} = 300 K) in Co$_2$MnGa:} Temperature dependence of electric (a) and thermal conductivity (b). (c) Magnetization M  at room temperature.  (d) Anomalous Hall effect (e) anomalous Nernst effect, and (f) anomalous thermal Hall effect at room temperature. In all cases in this paper, \textit{B} $\parallel$ [110], \textit{I} $\parallel$ [001]. The insets in (d-f) show the experimental configurations for Hall, Nernst and thermal Hall measurements.
\label{fig:longitude}}
\end{figure}

Figure \ref{fig:longitude}(a) and (b) show the temperature dependence of the thermal conductivity, $\kappa_{xx}$ and electric resistivity, $\rho_{xx}$. Resistivity varies from 125 $\mu \Omega$ cm at room temperature to 60 $\mu \Omega$ cm at low temperature.  This implies a short mean-free-path for electrons and given the magnitude of $\kappa_{xx}$ a dominant role for phonons (and eventually magnons) in longitudinal thermal transport.

As seen in Figure \ref{fig:longitude}(c), room temperature  magnetization  saturates to 4 $\mu_B$/f.u., similar to what was reported previously~\cite{Sakai2018,Manna2018}.  In contrast to other topological magnets~\cite{Li2017,Xu2018,Ding2019}, no clear hysteresis loop is visible, indicating that the domain walls smoothly propagate without pinning when the magnetic field is swept. The field dependence of the Hall, Nernst and Righi-Leduc effects at room temperature are shown in Fig.\ref{fig:longitude}(d-f). They all display a behavior similar to the magnetization. The presence of a large anomalous component is visible. The steep initial slope is replaced by a much smaller one at higher magnetic field. The anomalous component of the Hall resistivity ($\simeq 13 \mu \Omega$ cm) is slightly below what was previously reported ($\simeq 15 \mu \Omega$ cm)\cite{Sakai2018} and the anomalous Nernst signal ($\simeq 8 \mu$ V/ K) is somewhat larger than a previous report ($\simeq 6.5 \mu$  V/ K)~\cite{Sakai2018}. We note that the room-temperature anomalous $S^A_{xy}$ in crystals~\cite{Sakai2018,Guin2019} is three to four times larger than in thin films~\cite{Reichlova2018,Park2019}.

\begin{figure}
\includegraphics[width=8.5cm]{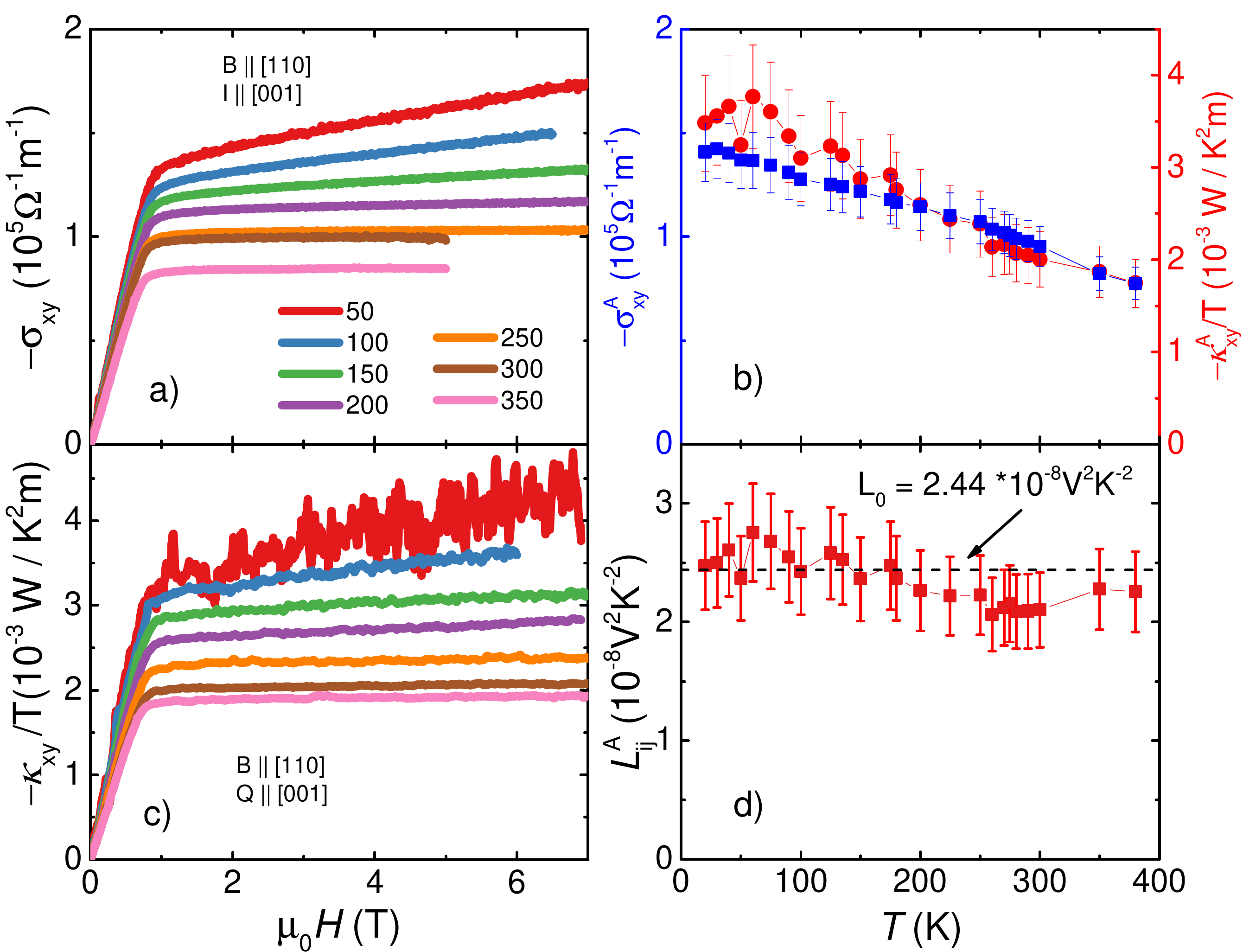}
\caption{\textbf{Temperature evolution of Hall conductivity and thermal Hall  conductivity and Anomalous transverse Lorenz number in Co$_2$MnGa:} (a) and (b) Field dependent Hall conductivity $\sigma_{xy}$ and transverse thermal conductivity divided by T $\kappa_{xy}/T$ at selected temperatures. (c) Temperature dependent anomalous Hall conductivity and transverse thermal conductivity. (d) The anomalous transverse Lorenz ratio ($L^A_{xy}$ = $\kappa^A_{xy}$/$\sigma^A_{xy}T$).}
\label{Fig:normal}
\end{figure}

Fig.\ref{Fig:normal}(a,c) shows the thermal evolution of the Hall conductivity ($\sigma_{xy}$) and the  thermal Hall conductivity  ($\kappa_{xy}/T$) in a field window extended to 7 T (The raw data can be seen in the supplement~\cite{Supplementary}). In both cases, given the clear change in the slope of the curve, the extraction of the anomalous components is straightforward. Fig.\ref{Fig:normal}(b) shows how the anomalous Hall conductivity, $\sigma^A_{xy}$, and the anomalous thermal Hall conductivity $\kappa^A_{xy}/T$ depend on temperature. Both increase almost twofold as the system is cooled from 300 K to 4 K.  The validity of the anomalous WF law for transverse response can be checked by comapring the anomalous Lorenz number $L^A_{xy} = \frac{\kappa^A_{xy}}{T\sigma^A_{xy}}$ with the Sommerfeld value $L_{0}= \frac{\pi^2}{3}(\frac{k_B}{e})^2$ where $k_B$ and $e$ are the Boltzmann constant and the elementary charge of electron. As seen in Fig.\ref{Fig:normal}(d), within experimental margin, we find that L remains close to $L_0$ and the anomalous WF law is verified.



\begin{figure}
\includegraphics[width=8.5cm]{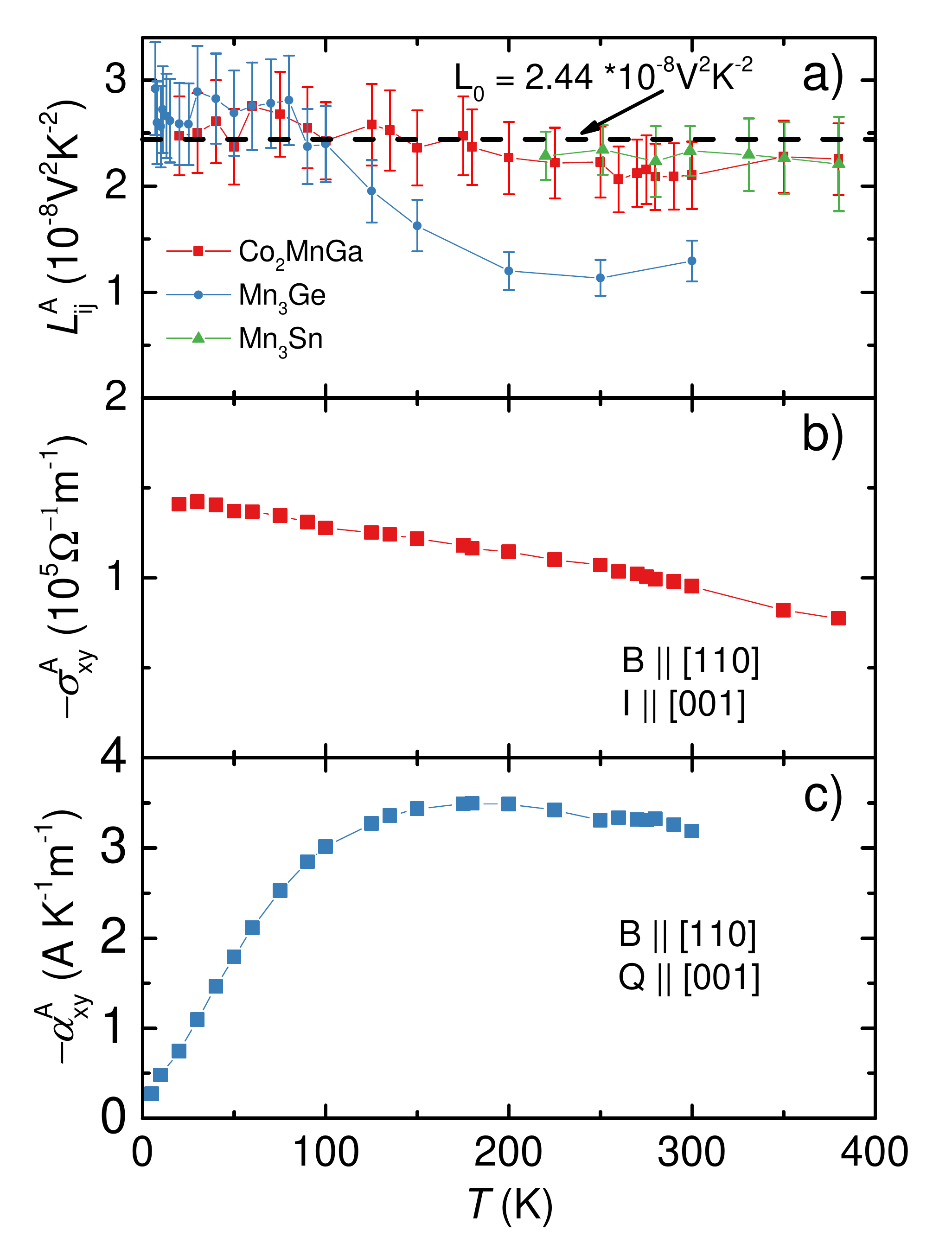}
\caption{\textbf{Anomalous Transverse coefficients:}  (a) The anomalous Lorenz number $L^{A}_{xy}$ = $\kappa^{A}_{xy}/T\sigma^{A}_{xy}$ remains close to the Sommerfeld value $L_0$. The anomalous Lorenz number of Mn$_3$Sn~\cite{Li2017} and Mn$_3$Ge~\cite{Xu2018} are also shown. The dashed horizontal line represents  L$_0$. (b) The anomalous Hall conductivity $\sigma^A_{xy}$ as a function of temperature. (c) The anomalous transverse thermoelectric conductivity, $\alpha^A_{xy}$, as a function of temperature. For comparison with data reported previously~\cite{Sakai2018,Guin2019}, see the supplement~\cite{Supplementary}.
\label{fig:Anom.WF}}
\end{figure}

In Fig.\ref{fig:Anom.WF}(a), we compare the evolution of  the anomalous Lorenz number $L^A_{ij}$ in Co$_2$MnGa with those found in two other topological magnets.  In the case of Mn$_{3}$Sn ~\cite{Li2017}, the anomalous WF law was found to be held above 150 K.  The commensurate  triangular magnetic order is taken over by a helical one below this temperature. In Mn$_3$Ge~\cite{Xu2018}, such a transition is absent and thus the validity of the anomalous WF law could be checked down to cryogenic temperatures, but a sizeable difference between $L^A_{xy}$ and L$_0$ was observed above 100 K and attributed to a  mismatch between thermal and electrical summations of the Berry curvature~\cite{Xu2018}. In common elemental ferromagnets, like Ni~\cite{Onose2008} and Fe~\cite{Li2017},  $L^A_{xy}$ was found to be close to L$_0$ at low temperature and a downward deviation emerges on heating, presumably due to inelastic scattering ~\cite{Li2017}. Thus, Co$_2$MnGa is the first case of a  magnet in which the anomalous WF law remains valid between 4 K and 300 K. This means that not only inelastic scattering is irrelevant, but also the thermal and electrical summations of the Berry curvature match each other. Thus, the variation of the Berry curvature near the Fermi energy is not abrupt in contrast to the case of Mn$_3$Ge~\cite{Xu2018}.

Fig.\ref{fig:Anom.WF}(b) and (c) show the temperature dependence of $\sigma^A_{xy}$ and  $\alpha^A_{xy}$. The magnitude of $\alpha^A_{xy}$ (which attains 1400 $\Omega$ cm$^{-1}$ at 2 K and decreases to 1000 $\Omega$ cm$^{-1}$ at room temperature) is remarkably large and  in reasonable agreement with previous studies~\cite{Sakai2018,Guin2019}. The anomalous transverse thermoelectric conductivity, $\alpha^A_{xy}$ displays a non-monotonous temperature dependence decreasing only below 150 K and vanishing as expected in the zero-temperature limit. Our $\alpha^A_{xy}$ data is in reasonable agreement with Sakai and co-workers' data ~\cite{Sakai2018}, but does not match what was reported by Guin \textit{et al.}~\cite{Guin2019}, discussed in the supplement~\cite{Supplementary}, this is due to an elementary mistake in manipulating the sign of the two components of $\alpha^A_{xy}$. Rectifying this mistake, one finds that the three sets of data give comparable value at room-temperature~\cite{Supplementary}.

The low-temperature slope of $\alpha^A_{xy}$ can be quantified by plotting $\alpha^A_{xy}/T$ as a function of temperature ~\cite{Supplementary}. According to the Mott's relation, this slope quantifies the variation of the low-temperature $\sigma^A_{xy}$ caused by an infinitesimal shift in the chemical potential. The magnitude of the extracted slope in Co$_2$MnGa ($\alpha^A_{xy}/T  \simeq -0.037 AK^{-2}$m$^{-1}$) is smaller than what was seen in Co$_3$Sn$_2$S$_2$~\cite{Ding2019}. In the latter case, the extracted slope was found to be in good agreement with the observed variation of low-temperature $\sigma^A_{xy}$ caused by a small shift in the chemical potential~\cite{Ding2019}. Our data reveals a well-defined $\alpha^A_{xy}/T$ and merges to the -ln T behavior above 100 K, similarly to the one reported by a previous study  ~\cite{Sakai2018}.(See the discussion in ~\cite{Supplementary}).


\begin{figure}
\centering
\includegraphics[width=8cm]{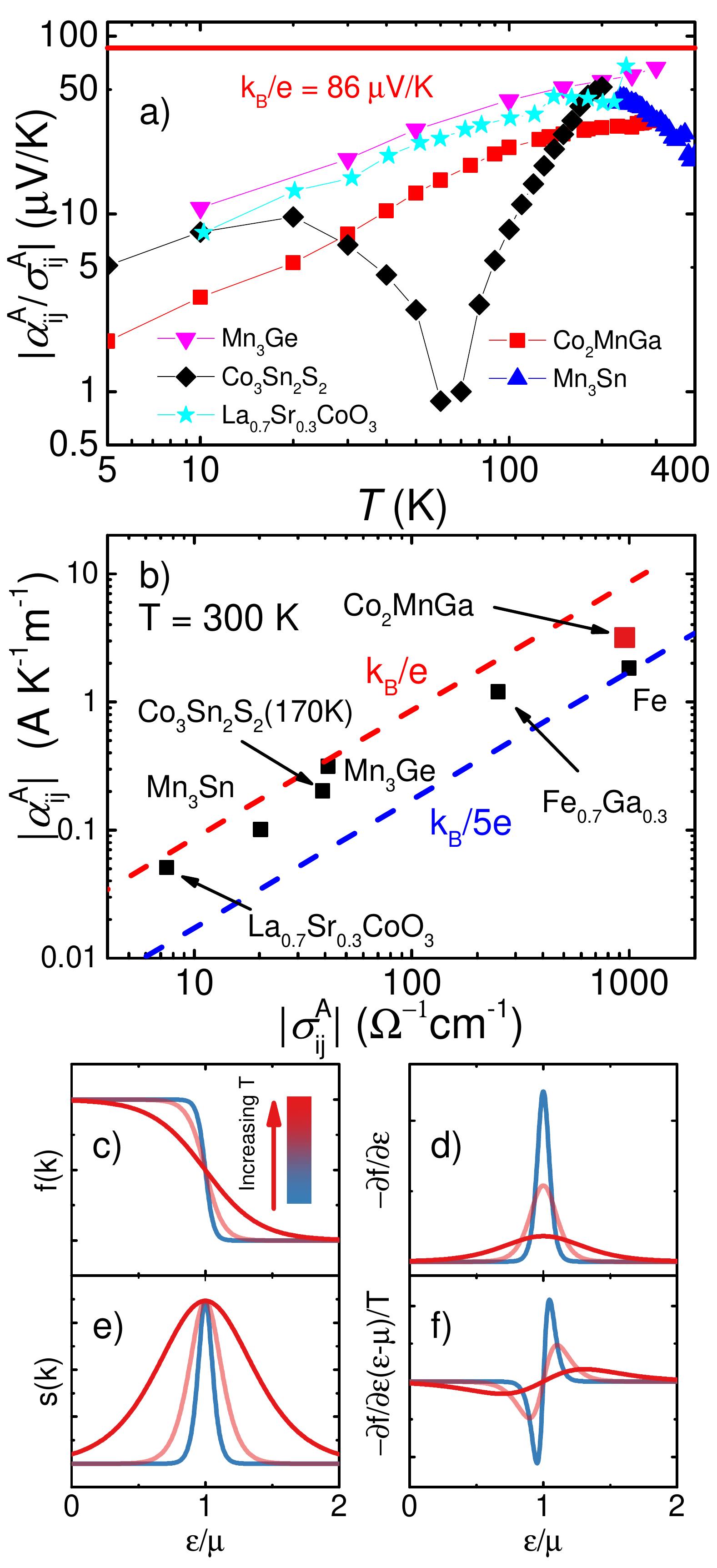}
\caption{\textbf{The  universal relation found in  $\alpha^A_{ij}/\sigma^A_{ij}$ ratio and the pondering functions:} a): The $\alpha^A_{ij}/\sigma^A_{ij}$ ratio as a function of temperature in different magnets including Mn$_3$Sn\cite{Li2017},  Fe\cite{Li2017}, Mn$_3$Ge\cite{Xu2018}, La$_{0.7}$Sr$_{0.3}$CoO$_3$\cite{Miyasato2007} and Co$_3$Sn$_2$S$_2$\cite{Ding2019}). k$_B$/e is represented by a red solid line. b):  Room-temperature $\alpha^A_{ij}/\sigma^A_{ij}$ in different magnets(Fe$_{0.7}$Ge$_{0.3}$\cite{Nakayama2019}). All the points are shows at 300 K(Co$_3$Sn$_2$S$_2$ is taken at 170 K because of the magnetic order). They lie all between k$_B$/5e(blue line) and k$_B$/e(red line). Note the range of $\sigma^A$ between 8 $\Omega^{-1}$cm$^{-1}$ and 1000 $\Omega^{-1}$cm$^{-1}$.  c-f) Pondering functions in Fermi sea (c and e) and Fermi surface (d and f)  expressions of the anoamlous Hall coefficients. c) Fermi-Dirac distribution, $f(k)$, used for Fermi-sea $\sigma^A_{ij}$.  d) Its energy derivative,  used for Fermi-surface $\sigma^A_{ij}$. e) The entropy density of electrons, $s(k)$, used for Fermi-sea $\alpha^A_{ij}$. f) Its energy derivative, used for Fermi-surface $\alpha^A_{ij}$.
\label{fig:4}}
\end{figure}

We now turn our attention to the magnitude of the $\alpha^A_{ij}/\sigma^A_{ij}$ ratio. In Fig.\ref{fig:4}a), we present the temperature dependence of this ratio in  Co$_2$MnGa and compare it to the available data of other magnets. Among these five systems, Co$_3$Sn$_2$S$_2$ distinguishes itself by the sign change of $\alpha^A_{xy}$ around 60 K~\cite{Ding2019}. However, like the four other systems, this ratio tends towards saturation at a value which is a sizeable fraction of $k_B/e$.

As seen in Fig.\ref{fig:4}b, even though  $\sigma^A_{ij}$ changes by a factor of 100 among different magnetic materials, the room-temperature  $\alpha^A_{ij}/\sigma^A_{ij}$ ratio remains between k$_B$/5e and k$_B$/e, which is the natural units of the ratio of these two quantities. This observation begs an explanation within our present understanding of anomalous transverse coefficients caused by Berry curvature.

The following expressions link  the anomalous Hall and the anomalous thermoelectric conductivity to the Berry curvature $\Omega^z$

\begin{equation}\label{eq:sigma}
    \sigma_{xy}^A = \frac{e^2}{\hbar}\int_{BZ}\frac{d^3k}{(2\pi)^3}f(k)\Omega^z
\end{equation}

\begin{equation}\label{eq:alpha}
   \alpha_{xy}^A = \frac{ek_B}{\hbar}\int_{BZ}\frac{d^3k}{(2\pi)^3}s(k)\Omega^z 
\end{equation}

Here $f(k)$ is the Fermi-Dirac distribution and  $s(k)=-f(k)ln(f(k))-(1-f(k))ln(1-f(k))$ is the (von Neuman) entropy density of electron gas, which are plotted in Fig.\ref{fig:4}c and Fig.\ref{fig:4}e. These 'Fermi sea' expressions for $\sigma_{xy}^A$ and  $\alpha_{xy}^A$ have 'Fermi surface' counterparts where the pondering factors are $\partial f/\partial \epsilon$ and $\partial s/\partial\epsilon= \frac{\partial f}{\partial\epsilon}\frac{(\epsilon-\mu)}{k_BT}$ (See the supplement for a discussion on the equivalency between the two formalisms).

Eq.\ref{eq:sigma} implies that the anomalous Hall conductivity is an average of the Berry curvature over the occupied fermionic states. One can express this idea by writing~\cite{Ding2019}: $\sigma_{xy}^A \approx \frac{e^2}{\hbar} \frac{1}{c}<\frac{\Omega_B}{\lambda_F^2}>$, where $\lambda_F$ is the Fermi wavelength in the plane perpendicular to the magnetic field and c is the lattice parameter along the magnetic field. In contrast, Eq.\ref{eq:alpha} implies that $\alpha_{xy}^A$  averages the Berry curvature over the states, which have a finite entropy, which within a thermal thickness of the Fermi level. Therefore $\alpha_{xy}^A \approx \frac{ek_B}{\hbar}\frac{1}{c} <\frac{\Omega_B}{\Lambda^2}>$~\cite{Ding2019}, where $\Lambda = \sqrt{\frac{h^2}{2\pi m k_B T}}$ is the de Broglie thermal wavelength in the plane perpendicular to the magnetic field. According to these expressions,  $\alpha_{xy}^A$ vanishes in the low temperature limit, because $\Lambda$ will diverge. On the other hand, $\lambda_F$ and therefore $\sigma_{xy}^A$ are expected to be finite in the  whole temperature range. Now, the $\alpha_{xy}^A/\sigma_{xy}^A$  will be set by $\frac{k_B}{e}<\frac{\lambda_F^2}{\Lambda^2}>$. Therefore, as the system is warmed up, $\lambda_F$ and  $\Lambda$ become comparable in size, the ratio should approach k$_B$/e. This provides a simple explanation for the strong correlation between the magnitudes of $\sigma^A_{ij}(300K)$ and $\alpha^A_{ij}(300K)$ in all known topological magnetic systems.

Let us note that a similar approach, associating $\alpha^A_{ij}$ and $\Lambda$, explained why the magnitude of the anomalous Nernst effect ($S^A_{ij}$) in a given magnet anti-correlates with the mean-free-path~\cite{Ding2019}, while the ordinary Nernst effect correlates with the mean-free-path~\cite{Behnia2016}. In both cases, the driving idea is simple. The intrinsic anomalous  $\alpha^A_{ij}$ (like the intrinsic anomalous  $\sigma^A_{ij}$) should not depend on the mean-free-path, but on the average Berry curvature. In contrast, semiclassical $\alpha_{ij}$ (like seimclassical $\sigma_{ij}$) scale with the inverse of the square of the mean-free-path~\cite{Behnia2016}. Note also that the large anomalous thermoelectric response of Co$_2$MnGa is demystified by this approach.  The room temperature $\alpha^A_{xy}$ of 3 AK$^{-1}$m$^{-1}$ is large compared to other topological magnets with lower AHE. Interestingly, BCC iron, whose room-temperature AHE is only slightly lower ($\simeq$ 1000 $\Omega^{-1}$cm$^{-1}$~\cite{Dheer1967,Li2017}  has a room-temperature $\alpha^A_{xy}$ as large as 2 AK$^{-1}$m$^{-1}$~\cite{Li2017}. As the room-temperature $S^A_{xy}$ is 20 times larger in Co$_2$MnGa than in Fe, it is partly because of its larger room-temperature resistivity (125$\mu \Omega$cm in Co$_2$MnGa compared to 10$\mu \Omega$cm in Fe).

In summary, we studied the anomalous off-diagonal coefficients of Co$_2$MnGa and checked the validity of the anomalous transverse Wiedemann-Franz law in the whole temperature range. This confirms that the anomalous transverse response is fundamentally a Fermi surface property~\cite{Haldane2004}. We quantified  $\alpha^A_{xy}$  and found that in all known topological magnets its magnitude at room temperature has  a universal relation with the size of the anomalous Hall conductivity and proposed an explanation for this observation.

\textbf{Acknowledgements-} This work was supported by the National Science Foundation of China (Grants No. 51861135104 and No. 11574097), by  Agence Nationale de la Recherche (ANR-18-CE92-0020-01) in France, by CREST(JPMJCR18T3), Japan Science and Technology Agency, by Grants-in-Aids for Scientific Research on Innovative Areas (15H05882 and 15H05883) from the Ministry of Education, Culture, Sports, Science, and Technology of Japan, and by Grants-in-Aid for Scientific Research (19H00650) from the Japanese Society for the Promotion of Science (JSPS). Z. Z. was supported by the 1000 Youth Talents Plan and K. B. was supported by China High-end foreign expert program. L. X. acknowledges a PhD scholarship by the China Scholarship Council(CSC).

\noindent
* \verb|zengwei.zhu@hust.edu.cn|\\
* \verb|kamran.behnia@espci.fr|\\

\bibliography{Co2MnGa}
\bibliographystyle{nature}

\clearpage
\renewcommand{\thesection}{S\arabic{section}}
\renewcommand{\thetable}{S\arabic{table}}
\renewcommand{\thefigure}{S\arabic{figure}}
\renewcommand{\theequation}{S\arabic{equation}}

\setcounter{section}{0}
\setcounter{figure}{0}
\setcounter{table}{0}
\setcounter{equation}{0}

{\large\bf Supplemental Material for ''Anomalous transverse response of Co$_2$MnGa and universality of the room-temperature $\alpha^A_{ij}/\sigma^A_{ij}$ ratio across topological magnets''}

\section{Sample prepration and measurements}
The single crystal Co$_2$MnGa were prepared by the Czochralski method\cite{Sakai2018}. The resistivity and Hall measurements were performed in a commercial measurement system(PPMS, Quantum Design) using the standard four-probe method with a pair of current source(Keithley 6221) and DC-Nanovoltmeter(Keithley 2182)\cite{Xu2018}.
The thermal gradient was measured by Type-E thermocouple\cite{Xu2018} in PPMS with high vacuum environment. The Nernst(Seebeck) voltage was detected by a DC-Nanovoltmeter separately. For thermal Hall effect measurements, we set $\mu_{0}H$ at +(-) 1 T at different temperatures to get the anomalous thermal Hall conductivity. We also measured the  transverse thermal conductivity with the function of field at some selected temperatures to get the normal part of thermal Hall effect. We used those convention in the main text(See Fig.\ref{fig:longitude}) to determine the sign for the transverse signal(Hall, Nernst and Righi-Leduc).

\section{Carrier density and mobility}
\begin{figure}
\includegraphics[width=7.5cm]{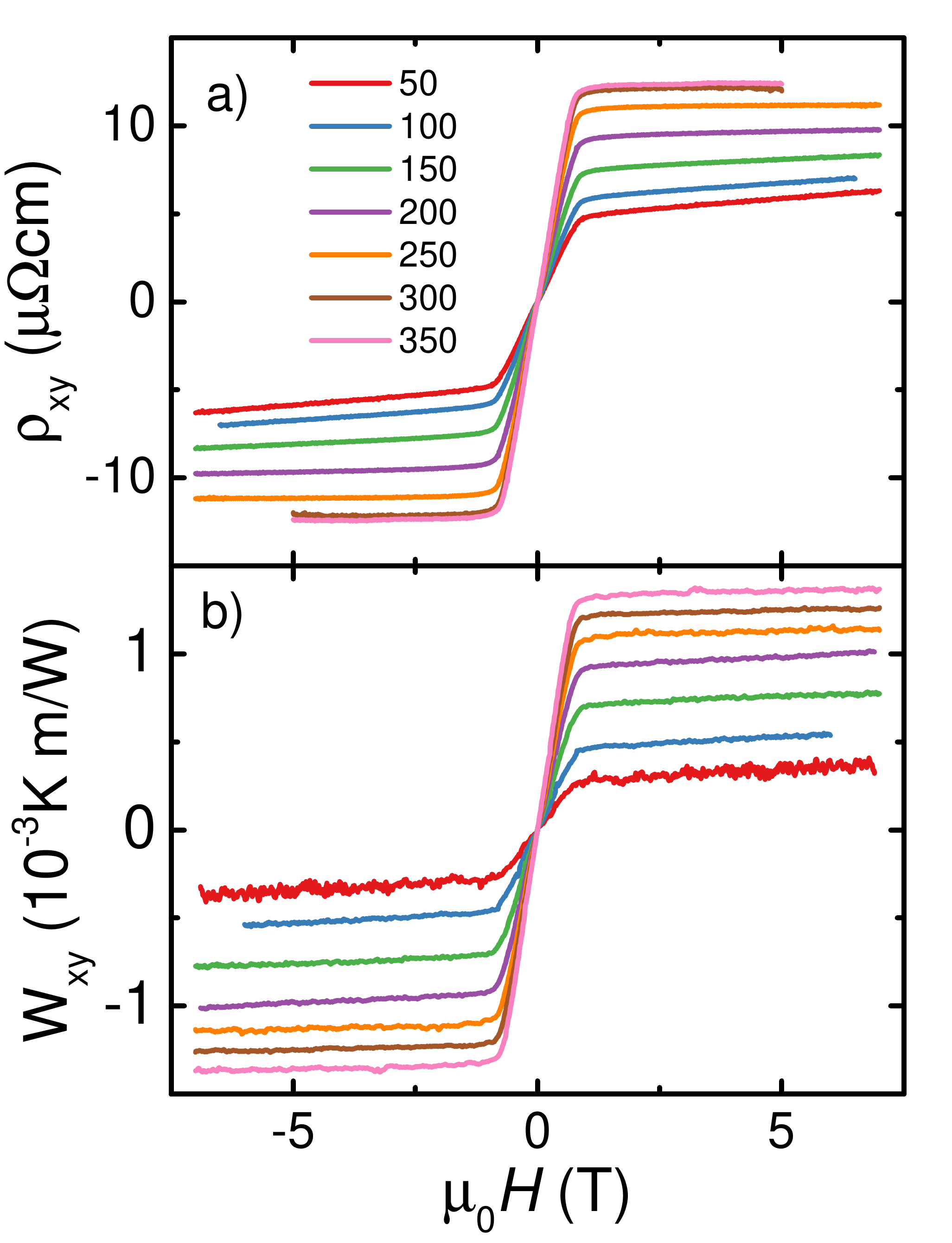}
\caption{The raw data: Temperature dependence of Hall resisitivity(a)) and thermal Hall ressitivity (b)) .
\label{fig:rho(B)-T}}
\end{figure}

In the low temperature(2 K), carrier density n is $\approx$ 2.8 $\times$ 10$^{21}$ cm$^{-3}$, using the single-band model(n = 1/\textit{e}R$_H$, where R$_H$ is the Hall coefficient), along with the longitude resistivity $\rho_{xx}$ = 5.8 $\times$ 10$^{-7}$ $\Omega$m, the carrier mobility results $\mu$ = 1/n\textit{e}$\rho$ $\approx$ 38 cm$^2$V$^{-1}$s$^{-1}$. Fig.\ref{fig:rho(B)-T} shows the temperature evolution of Hall effect and thermal Hall effect. As it's depicted in Fig.\ref{fig:rho(B)-T}a), the slope of ordinal Hall effect decrease a lot while increasing the temperature. We notes that it seems like the sign would change at a higher temperature.

\begin{figure}
\includegraphics[width=7.5cm]{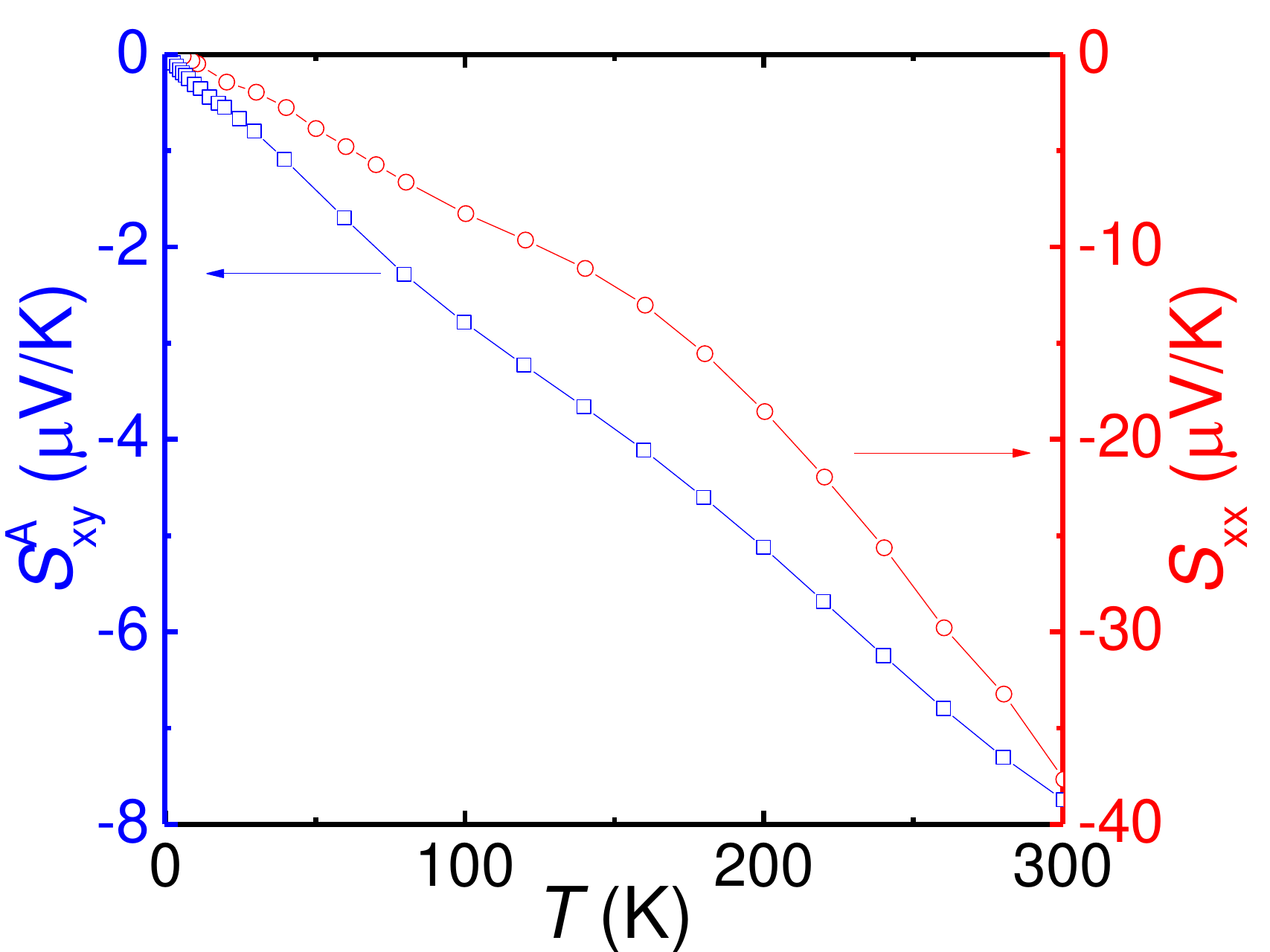}
\caption{\textbf{Anomalous Nernst coefficient and the Seebeck coefficient as a function of temperature:} The anomalous Nernst coefficient $S^A_{xy}$ extracted from the field dependence of the Nernst effect at different temperatures. It attains -7.8 $\mu$ V/K at room temperature. The temperature dependence of the Seebeck coefficient is also shown.
\label{fig:S1}}
\end{figure}

\section{The two components of $\alpha^A_{ij}$}
The charge density current in a solid depends on both the electric field $\textbf{\textit{E}}$ and the thermal gradient$\nabla \textbf{\textit{T}}$\cite{Behnia2015}:
\begin{equation}\label{eq:thermoelectric}
    \textbf{\textit{J}}_\textbf{e} =\vec{\sigma} \textbf{E}-\vec{\alpha} \nabla \textbf{T}
\end{equation}
In absence of  charge current, one has : $\textbf{\textit{J}}_e$ = 0 and therefore
$\vec{\sigma} \textbf{\textit{E}}-\vec{\alpha} \nabla \textbf{\textit{T}}$ = 0. Since  $S_{ij} = E_i/\nabla_jT$, one can write:

\begin{equation}
    \vec{\alpha}=\vec{\sigma} \vec{S}
\end{equation}
 Eq.\ref{eq:alpha} in the matrix form becomes:
\begin{equation*}
\begin{aligned}
    \begin{pmatrix}
      \alpha_{xx} & \alpha_{xy} \\
      \alpha_{yx} & \alpha_{yy} \\
    \end{pmatrix}
    =
    \begin{pmatrix}
      \sigma_{xx} & \sigma_{xy} \\
      \sigma_{yx} & \sigma_{yy} \\
    \end{pmatrix}
    \begin{pmatrix}
      S_{xx} & S_{xy} \\
      S_{yx} & S_{yy} \\
    \end{pmatrix}
\\
    =\begin{pmatrix}
      \sigma_{xx}S_{xx}+\sigma_{xy}S_{yx} & \sigma_{xx}S_{xy}+\sigma_{xy}S_{yy} \\
      \sigma_{yx}S_{xx}+\sigma_{yy}S_{yx} & \sigma_{yx}S_{xy}+\sigma_{yy}S_{yy} \\
    \end{pmatrix}
    \end{aligned}
\end{equation*}

As a consequence:

\begin{equation}\label{eq:ratio_alpha_sigma}
    \alpha_{xy} = \sigma_{xx}S_{xy} + \sigma_{xy}S_{yy}
\end{equation}

This leads to the following expression between $\alpha_{xy}$ and experimentally measurable quantities:

\begin{equation}\label{eq:ratio_alpha_sigma1}
    \alpha_{xy} = \frac{\rho_{yy}S_{xy} - \rho_{xy}S_{yy}}{\rho_{xx}\rho_{yy}-\rho_{xy}\rho_{yx}}
\end{equation}

Two different conventions for defining $S_{xy}$ have been used: $S_{xy} = E_x/\nabla_{y}T$ or $S_{xy} = E_x/-\nabla_{y}T$. The choice of convention for  $S_{xy}$ does not matter by itself. However, it can lead to mistakes on determining the magnitude of $\alpha_{xy}$.

\begin{figure}
\includegraphics[width=8.0cm]{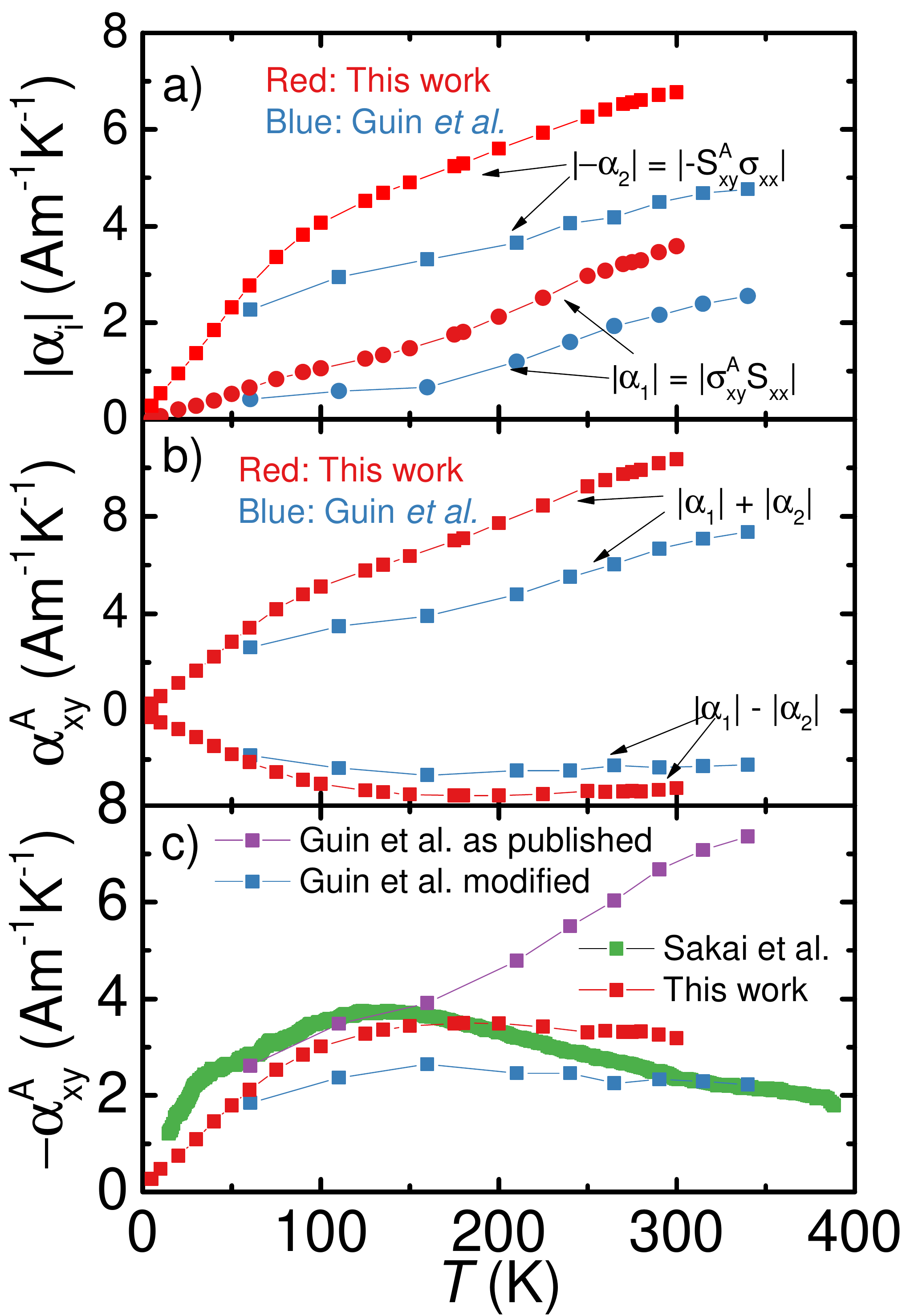}
\caption{\textbf{Two components of $\alpha^{A}_{xy}$ and the consistency} a) Data by Guin \textit{et al.}(blue) \cite{Guin2019} and this work(red) for the two components of $\alpha^{A}_{xy}$ in Eq.\ref{eq:ratio_alpha_sigma}. $\alpha_1$ = $\sigma^{A}_{xy}S_{xx}$ and $\alpha_2$ = -S$^{A}_{xy}\sigma_{xx}$. b) The sum and the difference of the two terms. Guin \textit{et al.} \cite{Guin2019} added the two components, while the correct procedure is to subtract them. We note that the S$^{A}_{xy}$ is positive in Guin \textit{et al.} but  negative here as a matter of convention. c) $\alpha^{A}$ of Co$_2$MnGa according to Guin \textit{et al.}(modified), Sakai \textit{et al.} and the present data. An overall consistency can be observed after correcting the mistake and assuming an opposite convention.
\label{fig:alpha_1_2}}
\end{figure}

\begin{figure}
\includegraphics[width=8.0cm]{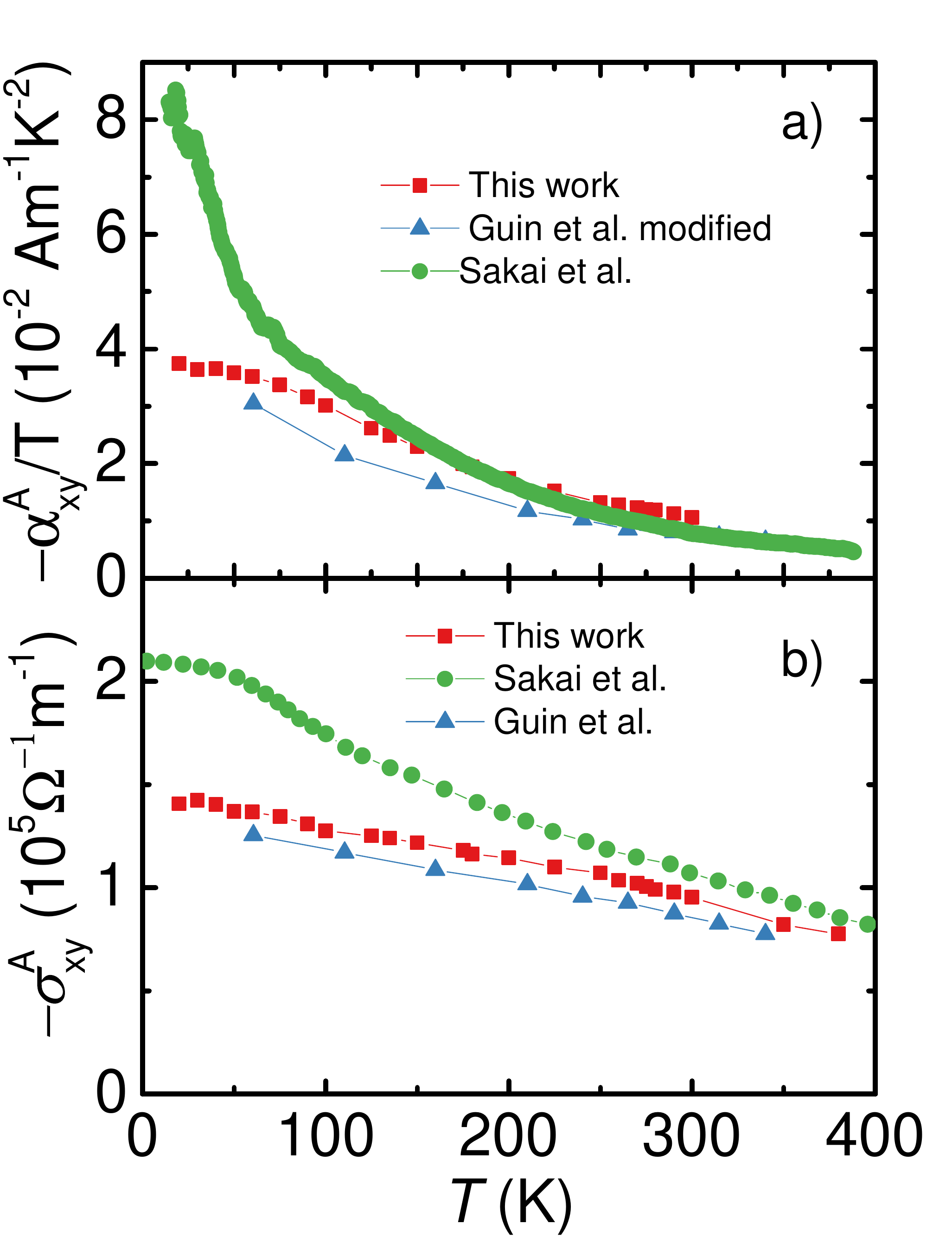}
\caption{\textbf{ Comparison of $\alpha^{A}_{xy}/T$ and $\sigma^{A}_{xy}$ reported in Co$_2$MnGa} a) $\alpha^{A}_{xy}/T$ according to Guin \textit{et al.}(with opposite sign), Sakai \textit{et al.} and this work. At low temperature the slope of $\alpha^{A}_{xy}/T$ is unambiguously found to be close to -0.037Ak$^{-2}$m$^{-1}$ in our data.  The slope in the data reported by Sakai \textit{et al.} is more ambiguous, but not very different.  b) The temperature dependence of  $\sigma^{A}_{xy}$ according to the thtee different groups. The data by  by Sakai \textit{et al.} is comparable to ours at room temperature, but slightly larger at low temperatures. The difference is attributable to a small change in the stoichiometry.
\label{fig:S4}}
\end{figure}

\begin{figure}
\includegraphics[width=8.0cm]{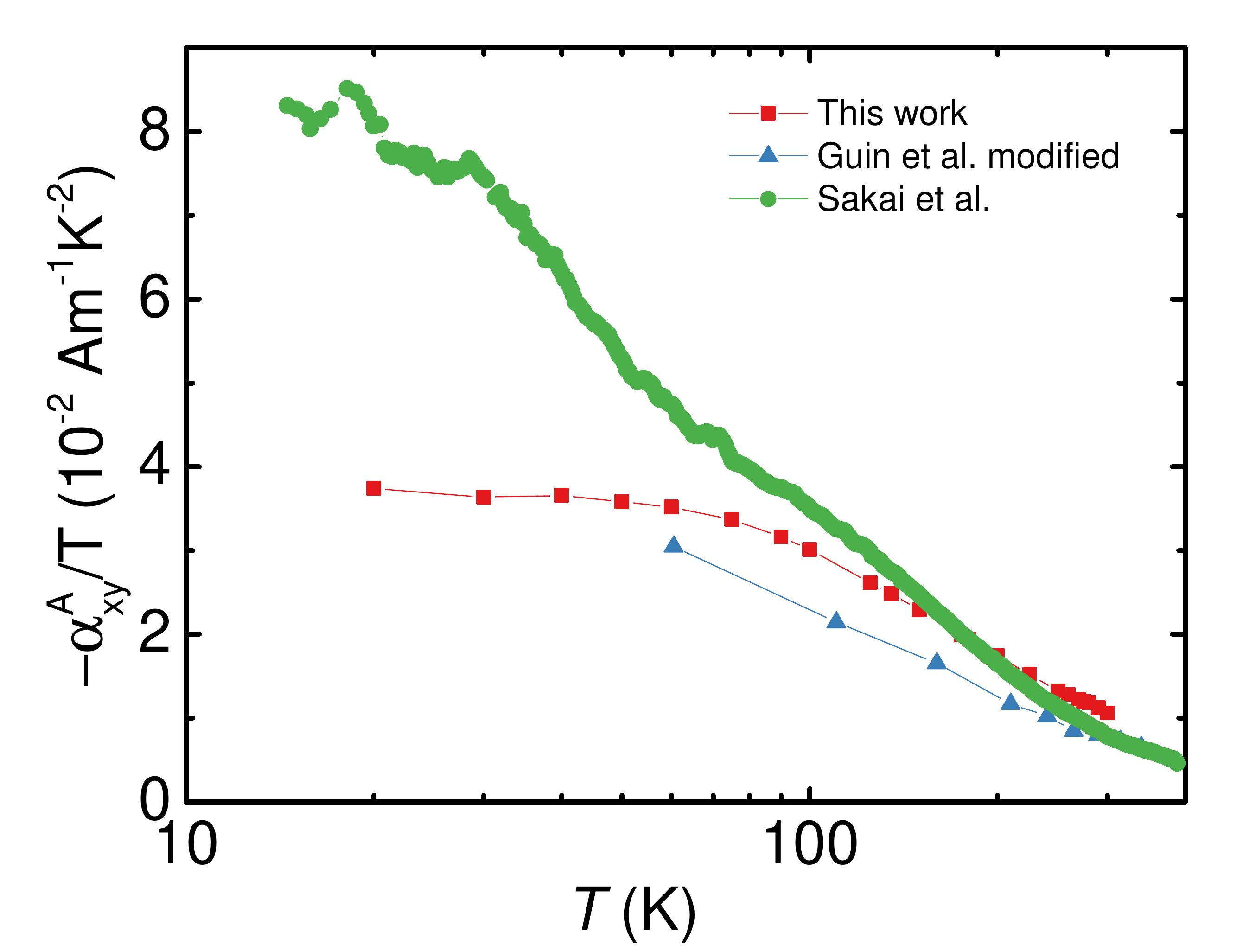}
\caption{\textbf{Logarithmic T dependence and quantum criticality-} In a semi-log plot,  $\alpha_{xy}/T$ is indeed linear in -$ln(T)$ in a finite temperature window, attributed to quantum criticality~\cite{Sakai2018}. In the zero-temperature limit, $\alpha_{xy}/T$ saturates to a finite value.
\label{fig:S5}}
\end{figure}

 \section{$\alpha^A_{xy}$ and $\sigma^{A}_{xy}$ according to different groups}

Guin \textit{et al.}\cite{Guin2019} report that $\alpha^A_{xy}$ in Co$_2$MnGa is as large as 7AK$^{-1}$m$^{-1}$. This is more than twice larger than what we or  Sakai \textit{et al.} find. The reason behind this discrepancy appears to be a simple mistake. They have apparently added instead of subtracting the  two components of $\alpha^A_{xy}$ in the right side of  Eq.\ref{eq:ratio_alpha_sigma}. As seen in Fig.\ref{fig:alpha_1_2}, the two components, namely -$\sigma_{xx}S^A_{xy}$ and $\sigma^A_{xy}S_{yy}$ have comparable values. Therefore the correct sign of these two  parts is crucial in setting $\alpha^{A}_{xy}$. After rectifying their data ( Fig.\ref{fig:alpha_1_2}), one sees that the correct $\alpha^A_{xy}$ from their measurements is non-monotonous like what was found by us and by Sakai \textit{et al.}(Fig.\ref{fig:alpha_1_2}c). The room-temperature magnitude of $\alpha^{A}_{xy}$ is close to 3 AK$^{-1}$m$^{-1}$ in the three sets of data.

Thus, the magnitude of room-temperature $\alpha^A_{xy}$ and $\sigma^{A}_{xy}$ in Co$_2$MnGa appears to be known beyond reasonable doubt. On the other hand, as one can see in Fig.\ref{fig:S4}, in the low -temperature limit, the data by Sakai \textit{et al.} indicates a non-monotonous $\sigma^{A}_{xy}$. The magnitude of $\alpha_{xy}/T$ in the zero-temperature limit according to their data is not very different from ours.

By plotting $\alpha_{xy}/T$ as a function of ln(T) (Fig.\ref{fig:S5}), one can see how the two sets of data comapre with each other. According to Sakai \textit{et al.}~\cite{Sakai2018}, $\alpha_{xy}/T$ is almost linear in ln(T) between  20 K and 200 K. A crossover to a constant behavior from ln T dependence appears below 20 K. In our data the temperature window in which $\alpha_{xy}/T$ is linear in ln(T) is restricted to a high-temperature range above 100 K and the saturation to a constant value below 100 K is very clear. The difference  in $\alpha_{xy}/T$ seen in the two sets of data may be ascribed to a small change in the stoichiometry and a  slight shift in the Fermi energy. Future studies will tell if there is a critical doping at which the $\alpha_{xy}/T$ remains linear in ln(T) down to the lowest measured temperature as expected in a quantum-critical scenario.

\section{Fermi-sea and Fermi-surface expressions for anomalous transverse coefficients}




Xiao and co-workers wrote the following expression for the anomalous off-diagonal thermoelectric conductivity:~\cite{Xiao2006}:
\begin{equation}\label{eq:alpha_Berry}
    \alpha^A_{xy} = \frac{e}{T\hbar}\int\frac{d^3k}{(2\pi)^3}\Omega^{z}(k)F(k)
\end{equation}

Here,  $F(k) = (\epsilon-\mu)f(k)+\frac{1}{\beta} ln[1+e^{-\beta(\epsilon-\mu)}]$, $\beta=1/k_{B}T$, $f(k)$ is the Fermi-Dirac distribution and $T$ is the temperature.

A variable change $(\epsilon-\mu)$ $\rightarrow$ $f(k)$~\cite{Bergman2010,Zhang2008} can show that $F(k)$ can be expressed in terms of the entropy density function $s(k)=-f(k)ln(f(k))-(1-f(k))ln(1-f(k))$.

\begin{equation*}\label{eq:ratio_alpha_sigma1}
    \begin{aligned}
    F(k)
    &=(\epsilon-\mu)f(k)+k_{B}Tln[1+e^{-\beta(\epsilon-\mu)}]\\
    &=k_{B}T\{\beta(\epsilon-\mu)f(k)+ln[1+e^{-\beta(\epsilon-\mu)}]\}\\
    &=k_{B}T\{ln(\frac{1}{f(k)}-1)f(k)+ln[1+\frac{1}{1/f(k)-1}]\}\\
    &=k_{B}T\{ln(\frac{1-f(k)}{f(k)})f(k)+ln[1+\frac{f(k)}{1-f(k)}]\}\\
    &=k_{B}T\{[ln(1-f(k))-ln(f(k))]f(k)-ln(1-f(k))\}\\
    &=k_{B}T\{-f(k)ln(f(k))-(1-f(k))ln(1-f(k))\}\\
    &=k_{B}Ts(k)
    \end{aligned}
\end{equation*}

The anomalous thermoelectric conductivity becomes:

\begin{equation}\label{eq:alpha_Berry}
    \alpha^A_{xy} = \frac{k_{B}e}{\hbar}\int\frac{d^3k}{(2\pi)^3}\Omega^{z}(k)s(k)
\end{equation}

This expression, used in the main text, is a Fermi-sea expression. The connection to the Fermi-sea expression can be seen by taking the derivative of $s(k)$

\begin{equation}\label{eq:alpha_Berry}
    \begin{aligned}
    \frac{\partial s}{\partial\epsilon}
    &= -(ln(f(k))+1)\frac{\partial f}{\partial\epsilon} -(-ln(1-f(k))-1)\frac{\partial f}{\partial\epsilon}\\
    &= (-ln(f(k))+ln(1-f(k)))\frac{\partial f}{\partial\epsilon}\\
    &= ln(\frac{(1-f(k))}{f(k)})\frac{\partial f}{\partial\epsilon}\\
    &= \frac{\partial f}{\partial\epsilon}\frac{(\epsilon-\mu)}{k_{B}T}\\
    \end{aligned}
\end{equation}

As in the case of anomalous Hall effect, the pondering factor in the Fermi-surface  expression, is the negative energy deviation of that in Fermi sea picture, $f$.
\end{document}